\documentclass[aps,prl,preprint,superscriptaddress,showpacs]{revtex4}
\usepackage{graphicx}
\usepackage{dcolumn}
\usepackage{bm}
\usepackage{color}
\begin{document}
\title{Carbon-doped ZnO: A New Class of Room Temperature Dilute Magnetic Semiconductor}
\author{H. Pan}
\affiliation{Department of Physics, National University of
Singapore, 2 Science Drive 3, Singapore 117542}
\author{J. B. Yi}
\affiliation{Department of Materials Science and Engineering, National University of
Singapore, 9 Engineering Drive 1, Singapore 117576}
\author{J. Y. Lin}
\affiliation{Department of Physics,
National University of Singapore, 2 Science Drive 3, Singapore
117542}
\affiliation{Institute of Chemical and Engineering Sciences, 
1 Pesek Road, Jurong Island, Singapore 627833}
\author{Y. P. Feng}
\email{phyfyp@nus.edu.sg} \affiliation{Department of Physics,
National University of Singapore, 2 Science Drive 3, Singapore
117542}
\author{J. Ding}
\email{msedingj@nus.edu.sg} \affiliation{Department of Materials Science and Engineering, National University of
Singapore, 9 Engineering Drive 1, Singapore 117576}
\author{L. H. Van}
\affiliation{Department of Materials Science and Engineering, National University of
Singapore, 9 Engineering Drive 1, Singapore 117576}
\author{J. H. Yin}
\affiliation{Department of Materials Science and Engineering, National University of
Singapore, 9 Engineering Drive 1, Singapore 117576}

\date{\today}

\begin{abstract}

We report magnetism in carbon doped ZnO. Our first-principles calculations based on 
density functional theory predicted that carbon substitution for oxygen in ZnO 
results in a magnetic moment of 1.78 $\mu_B$ per carbon.
The theoretical prediction was confirmed experimentally. 
C-doped ZnO films deposited by pulsed laser deposition with various carbon concentrations 
showed ferromagnetism with Curie temperatures higher than 400 K, and the measured 
magnetic moment based on the content of carbide in the films ($1.5 - 3.0 \mu_B$ per carbon) is
in agreement with the theoretical prediction. The magnetism is due to bonding 
coupling between Zn ions and doped C atoms. Results of magneto-resistance and abnormal Hall effect 
show that the doped films are $n$-type semiconductors with intrinsic ferromagnetism.
The carbon doped ZnO could be a promising room temperature dilute magnetic semiconductor 
(DMS) and our work demonstrates possiblity of produing DMS with non-metal doping.

\end{abstract}

\pacs{75.50.Pp, 
71.55.Gs, 
75.50.-y, 
85.75.-d 
}
\maketitle

Dilute magnetic semiconductor (DMS) with Curie temperatures ($T_c$) at or above 
room temperature is essential for practical spintronics applications. However, 
synthesis of such materials has been an experimental challenge. 
DMS is usually produced by doping semiconductors with transition metals (TMs). 
ZnO and GaN were theoretically predicted to be ideal candidates for room 
temperature DMS\cite{Dietl}.  
Even though ferromagnetism has been observed in a number of systems, 
experimental studies on TM doped ZnO have produced inconsistent results and 
the mechanism of ferromagnetism in TM doped ZnO remains unclear. 
It is speculated that TM dopants in ZnO form clusters or secondary phases, 
which are detrimental to applications of DMS. 
This promoted search for DMS based on alternative dopants. 
If non-TM dopants can be incorporated into ZnO and 
induce maganetism, DMS thus produced would not suffer from problems related
to precipitates of dopants since they do not contribute to ferromagnetism. 
For example, copper doping in ZnO and GaN have been investigated and it has
been confirmed experimentally that both Cu doped ZnO and GaN are room
temperature DMS. \cite{Wu1,Park,Chien,Feng,Ye,Buchholz,Lee}
Other DMS obtained by doping with non-TMs, or without doping at 
all, were also reported, such as Mg doped AlN, Sc-doped ZnO and undoped HfO$_2$ films. 
\cite{Wu2,Coey,Venkatesan_PRL,Venkatesan_Nature}

Ferromagnetism was also reported in a number of carbon 
systems.\cite{Makarova,Kopelevich,Esquinazi,Wang,Kopelevich03,Likodimos,Rode,Han,Esquinazi03,Esquinazi04,Talapatra}
Some of these studies have speculated that intrinsic 
carbon defects could be responsible for the observed magnetic 
properties. Carbon adatoms on carbon nanotube\cite{Lehtinen} and carbon substitutional
doping in boron nitride nanotube\cite{Wu3} were predicted to induce magnestism in
the respective systems.  
It is therefore of interest to investigate the effect of carbon doping in ZnO 
and explore the possibility of using carbon as dopant to produce ZnO based DMS. 
Understanding the mechanism of ferromagnetism in non-metal doped ZnO is 
useful in exploring new areas of dilute magnetic semiconductors. 
In this letter, we present our computational and experimental studies on 
ZnO films doped with carbon and show that the carbon doped ZnO can be a 
promising DMS.

First-principles calculations based on the density functional theory (DFT) and 
the local spin density approximation (LSDA) was carried out to investigate 
carbon doping of ZnO. The plane-wave basis and pseudopotential approach,
\cite{Payne} as implemented in the CASTEP code, \cite{Milman} was used 
in our study. In the total energy calculations, the ionic potentials were 
described by the ultrasoft non-local pseudopotential proposed by Vanderbilt, 
\cite{Vanderbilt} and the exchange-correlation functional parameterized 
by Perdew and Zunger \cite{Perdew} was used. The system was modeled with 
a periodic supercell of $9.787\times 9.787 \times 10.411$ \AA$^3$ with 18 
formula units of wurzite ZnO, which is sufficient to avoid interaction 
of C atom with its images in neighboring supercells. An energy cut-off 
of 310 eV was used for the plane wave expansion of the electronic wave 
function. Special $k$ points were generated with a $4\times 4\times 3$ grid based on 
Monkhorst-Pack scheme.\cite{Monkhorst} Good convergence was obtained with these parameters. 
The total energy was converged to $2.0\times 10^{-5}$ eV/atom while the Hellman-Feynman 
force was smaller than $5.0\times 10^{-2}$ eV/\AA\ in the optimized structure. We consider 
three types of carbon doping: carbon interstitial (C$_{\rm I}$), carbon substitution 
at Zn site (C$_{\rm Zn}$) and carbon substitution at oxygen site (C$_{\rm O}$). The calculated 
lattice constants ($a=3.25$ \AA, $c=5.20$ \AA) for bulk ZnO are in good agreement with 
the experiment values ($a=3.25$ \AA, $c=5.21$ \AA).\cite{Madelung} The calculated band gap is 
1.13 eV within LSDA, which is consistent with the reported results.\cite{Kohan,Xu}

Our calculations predicted magnetism in carbon doped ZnO and revealed that 
it results from carbon substitution for oxygen. Figure~\ref{fig1} shows the 
calculated local density of states (LDOS) for the carbon dopant and the 
neighboring Zn atoms. Strong coupling between the carbon $s$ and $p$ orbitals 
and the $s$ orbital of Zn can be seen. The interaction causes the carbon 
$2s$ orbital around $-9$ eV and the carbon $2p$ orbital near 2.3 eV to split. 
The spin-up bands are fully occupied while the spin-down bands are partially 
filled, resulting in a magnetic moment of 1.78 $\mu_B$ per carbon dopant. 
The magnetic moment is mainly contributed by the carbon $p$ orbitals (0.80 $\mu_B$), 
while each of the neighboring Zn atoms and second nearest neighboring oxygen 
atoms also contribute a small part (0.1 $\mu_B$ and 0.04 $\mu_B$, respectively). 
The estimated formation energy of the C$_{\rm O}$ defect is 5.3 eV. Similar
calculations on carbon substitution for Zn and interstitial carbon showed that
they do not result in magnetism.

To verify the theoretical predition, we prepared C-doped ZnO films using pulsed-laser 
deposition (PLD).   
ZnO/C$_x$ targets with carbon concentrations $x = 0$, 0.5, 1, 5 and 10 
were prepared by sintering mixed ZnO (99.9\%) and carbon (99.9\%) powders in 
nitrogen atmosphere at 1273 K for 30 min. The base pressure was 
$<1\times 10^{-8}$ torr.
The C-doped ZnO films were deposited on sapphire (0001) substrate using a 
KrF excimer laser operating at 248 nm and a fluence of 1.8 J cm$^{-2}$. 
No ferromagnetism was found in the pristine ZnO, carbon powders and sintered 
ZnO/C targets, as well as the substrates. The films were deposited at 673 K 
and a vacuum better than $10^{-7}$ torr in order to ensure epitaxial growth 
and avoid carbon loss during deposition. The details of the sample preparation 
procedure was reported previously.\cite{Van} The carbon concentration was 
estimated based on the secondary ion mass spectrometry (SIMS) analysis with 
the support from X-ray photoelectron spectroscopy (XPS) and scanning electron
microscopy (SEM)
examinations. The film thickness was chosen to be 200 nm in order to ensure 
an accurate estimation of magnetization and to minimize the substrate effect. 
Three samples with target carbon concentrations of 0, 1 and 5 at\%,
designated as Sample A, Sample B and Sample C, respectively, were 
selected for the detailed structural and magnetic study as listed in 
Table~\ref{tab1}.

The structures of the ZnO films were characterized by X-ray diffraction (XRD) 
and SIMS.
Our XRD study indicated that both pure and C doped ZnO films show good epitaxy 
with the sapphire substrate, similar to that of previous study.\cite{Van} 
SIMS analysis showed that Zn and O are uniformly distributed in the C-doped 
films. Compared with SIMS spectrum of standard reference sample, the average 
carbon concentration was estimated to be 1 at\% and  2.5 at\% for the ZnO$+$1 
at\%C and ZnO$+$5 at\%C targets, respectively (Table~\ref{tab1}). It is noted
that the carbon concentration in the film is lower than that in the target and
thus the carbon loss is severe when the carbon concentration in the target is
a higher ($>3-4$ at\%).

Carbon can exist in different forms in the grown samples. To verify that 
carbon is doped into the ZnO films, we carried out further analysis on
the samples using XPS and Raman 
pectroscopy. Two groups of peaks in the C1s binding energy, namely a large 
peak at 284.6 eV and a few peaks in the energy range of $280-284$ eV were 
observed in the XPS analysis. The peak at 284.6 eV can be attributed to 
``free carbon'' (graphite and/or carbon from contamination). Our Raman 
spectroscopic study confirmed the presence of graphite. New carbon species 
with C1s binding energy between 280 and 284 eV observed in the C-doped ZnO 
films suggests presence of carbon atoms in the carbide form,\cite{Ramqvist} 
which is an indication of carbon substitution for oxygen
and formation of Zn-C bonds in the carbon doped ZnO films.

Magnetic properties of all three samples were investigated. As expected,
the pure ZnO film without C doping is non-magnetic, whereas 
both C doped ZnO films show ferromagnetism at room temperature. A detailed 
magnetic and electronic investigation was carried out for the two samples 
- Sample B with a relative low C-doping concentration and Sample C with a 
high C-doping concentration. The hysteresis loop of Sample B at 300 K, 
shown in the inset of Fig.~\ref{fig2}, clearly shows ferromagnetism.
The temperature dependence of magnetization shows a Brillouin type 
magnetization (Fig.~\ref{fig2}). The magnetization at 400 K for sample B 
is 3.8 emu/cm$^3$, indicating that the Curie temperature of the film is 
higher than 400 K. The exact Curie temperature could not be directly 
determined in our experiment due to the limitation of the SQUID system 
which has a temperature range of $\le 400$ K. However, we fitted the 
magnetization-temperature curve to the Bloch law,\cite{Theodoropoulou}
$1-M_s/M_0=BT^{3/2}$, where $M_s$ is the magnetization at temperature $T$, 
$M_0$ is the saturation magnetization of the film, and $B$ is a constant. 
The fitting yielded $B=5.73\times 10^{-5}$ K$^{2/3}$ and $M_0=7.26$ 
emu/cm$^3$. The Curie temperature of Sample B was thus estimated to be 
approximately 670 K. Similar fitting shows that Sample C has a higher 
Curie temperature, approximately 850 K.

We also investigated the dependence of the measured saturation 
magnetization on carbon concentration in the target, and the results 
are shown in Fig.~\ref{fig3}. The magnetization increases rapidly for 
low carbon concentrations ($0-1$ at\%), but becomes saturated when the 
carbon concentration in the target reaches about 5 at\%. Incidently, 
our SIMS and XPS analysis showed that the carbon concentration in the 
carbide form increases with the target concentration until about 5 at\%
beyond which only the amount of graphite carbon increases and the 
amount of carbide in the film remains fairly constant, indicating   
that free carbon do not contribute to the ferromagnetism.
By assuming that the XPS peak at 284.6 eV is completely attributed to 
graphite carbon, and that the carbon due to contamination is negligible,
we estimated the magnetic moment per carbon in the ZnO films and obtained values in the 
range of $2.5 - 3.0$ $\mu_B$ for Sample B and $1.5 - 2.5$ $\mu_B$ for 
Sample C, respectively (Table~\ref{tab1}). Magnetic moments per carbon 
of other samples are shown in the inset of Fig.~\ref{fig3}. It can be
concluded that the magnetic moment in the carbide form is 
between 1.5 and 3.0 $\mu_B$ per carbon for all the C-doped samples, 
which is in 
agreement with the theoretical prediction of 1.78 $\mu_B$.

As reported previously,\cite{Dietl} magneto-resistance (MR) and 
abnormal Hall effect (AHE) have been often observed in many DMS's. 
The AHE is a strong evidence of intrinsic ferromagnetism due to 
interactions between carriers and spins. In this work, we have studied 
AHE and MR of carbon-doped ZnO.  Figure~\ref{fig4} shows the Hall effect 
of Sample B at different temperatures. The total Hall effect can be 
expressed as $\rho_{xy}=R_0B+R_s\mu_0M$, where $B$ is magnetic induction, 
$\mu_0$ the magnetic permeability, and $M$ the magnetization. The first
term above ($R_0B$) represents the ordinary Hall effect, whereas the 
second term ($R_s\mu_0M$) denotes the 
abnormal Hall effect. The normal Hall effect indicates that the 
films are $n$-type semiconductor. The hysteresis loop of the Hall 
effect after the deduction of the normal Hall effect is shown in 
the inset. The shape of the Hall effect hysteresis loop is similar 
to that of the magnetic hysteresis loops measured by SQUID as shown 
in Fig.~\ref{fig2}. The Hall voltage strongly increases with the 
decreasing of temperature. The abnormal Hall effect also increases 
with the increasing C-doping concentration, as shown in Table~\ref{tab1} 
for Sample C. While Hall mobility decreases with the C-doping 
concentration, as shown in Table~\ref{tab1}. In addition, 
MR up to 0.5\% was present in C-doped ZnO. 
MR was proportional to the measured magnetization.

Ferromagnetism in TM-doped DMS can be due to either double exchange
or $p$-$d$ hybridization mechanism.\cite{akai,sato} In the latter,
the TM dopant hybridizes strongly with its neighboring anions of the 
host semiconductors, and the neighboring anions are spin polarized 
with magnetization in the same order of magnitude as that of the dopant 
and couple ferromagnetically to the dopant. 
Other dopants in turn couple to the spin polarized anions in the same 
way for an energy gain, resulting in an indirect FM coupling among 
dopants. A similar mechanism can be expected for ferromagnetism in
carbon doped ZnO, except that the hybridization here is between the $s$ and $p$ 
orbitals of the dopant and the $s$ orbitals of the neighboring zinc
atoms. As discussed above, the LDOS shown in Fig.~\ref{fig1} 
indicate a strong hybridization between carbon and its neighboring 
Zn atoms. This strong hybridization induces finite magnetization on 
carbon as well as the neighboring Zn atoms. However, further study
is necessary to clarify the mechanism for ferromagnetism in carbon
doped ZnO.

In conclusion, we demonstrated, both theoretically and experimentally,
that carbon can be ferromagnetic when substituting for oxygen in the ZnO 
environment. Carbon doped ZnO films show an intrinsic $n$-typed 
ferromagnetic behavior, with Curie temperatures well above room
temperature, which make carbon doped ZnO a promising room
temperature DMS and a potentially useful material for spintroncs 
devices. The ferromagnetism originates from the bonding between 
zinc and doped carbon. Ferromagnetism from carbon doped semiconductors
represents a new class of DMS and it opens new possibilities of producing 
DMS with non-metal doping.


\clearpage

\begin{table}
\caption {The properties of ZnO with and without carbon doping.}
\begin{ruledtabular}
\begin{tabular}{lccc}
	& Sample A	& Sample B	& Sample C \\ \hline
Carbon concentration of the target (at\%)	& 0	& 1		& 5\\
Measured C concentration (at\%)			& -	& $\sim 1$	& $\sim 2.5$\\
Magnetization at 300 K (emg/cm$^3$)		& 0	& 3.8		& 7.1\\
Magnetization at 5 K (emg/cm$^3$)		& 0	& 7.2		& 10.1\\
Estimated Curie temperature (K)			& 0	& $\sim 670$	& $\sim 850$\\
Moment per C in carbide state			& -	& 2.0-3.0	& 1.5-2.5\\
Resistivity ($\Omega\cdot$cm)			& 4	& 0.195		& 0.108\\
Hall mobility (cm$^2$/VS)			& 94.2	& 22.6		& 14.5\\
Carrier concentration ($10^{18}$/cm$^3$)	& 0.1	& 2.1		& 3.8\\
\end{tabular}
\end{ruledtabular}
\label{tab1}
\end{table}

\clearpage

\begin{figure}[h]
\caption{Calculated total (top panel) and local density of states 
for the carbon dopant and a neighboring Zn atom. The Fermi level 
is indicated by the dashed vertical line. } 
\label{fig1}
\end{figure}

\begin{figure}[h]
\centering
\caption{$M_s(T)/M_s({\rm 5K})$ versus temperature for Sample B and 
Sample C. The solid lines are curves corresponding to the Bloch law 
$1-M_s/M_0=BT^{3/2}$. The inset shows hysteresis loop taken 
at 300 K.}
\label{fig2}
\end{figure}

\begin{figure}[h]
\centering
\caption{Room temperature saturation magnetization as a function of 
the target carbon concentration. The inset shows the magnetic moment 
of carbon in the carbide state in the ZnO films as a function of the 
target carbon concentration.}
\label{fig3}
\end{figure}

\begin{figure}
\centering
\caption{The Hall voltage as a function of magnetic field for Sample B 
at different temperature. The inset shows the Hall effect curve at 10 K 
after the normal part is removed.}
\label{fig4}
\end{figure}

\clearpage
\begin{center}
\includegraphics{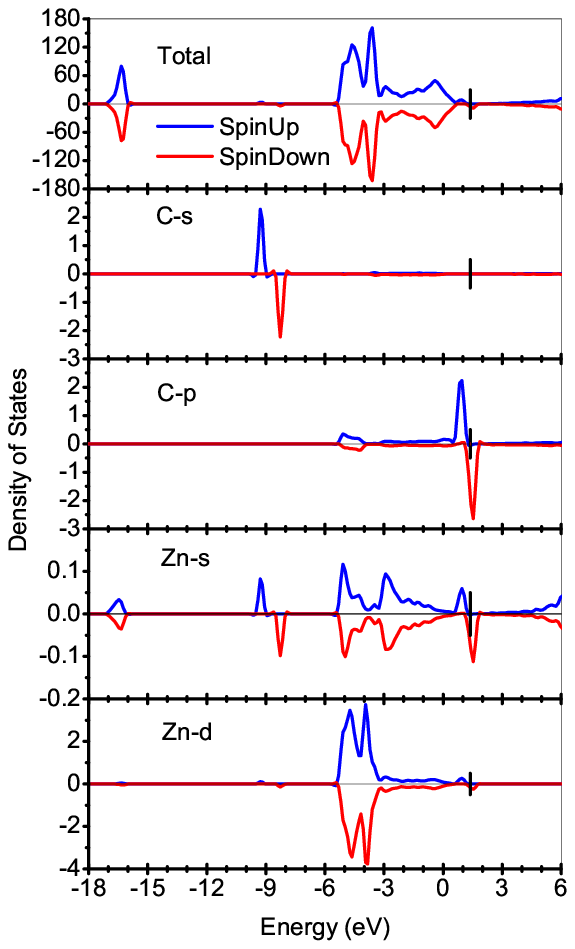} \\
Fig. 1 Pan et al.
\end{center}

\clearpage
\begin{center}
\includegraphics{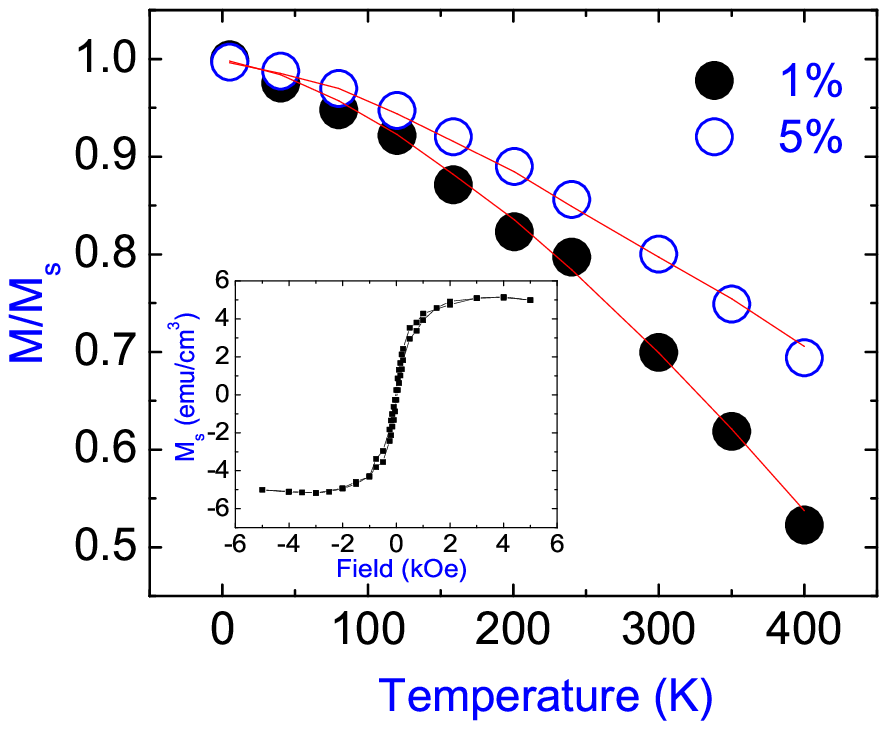} \\
Fig. 2 Pan et al.
\end{center}

\clearpage
\begin{center}
\includegraphics{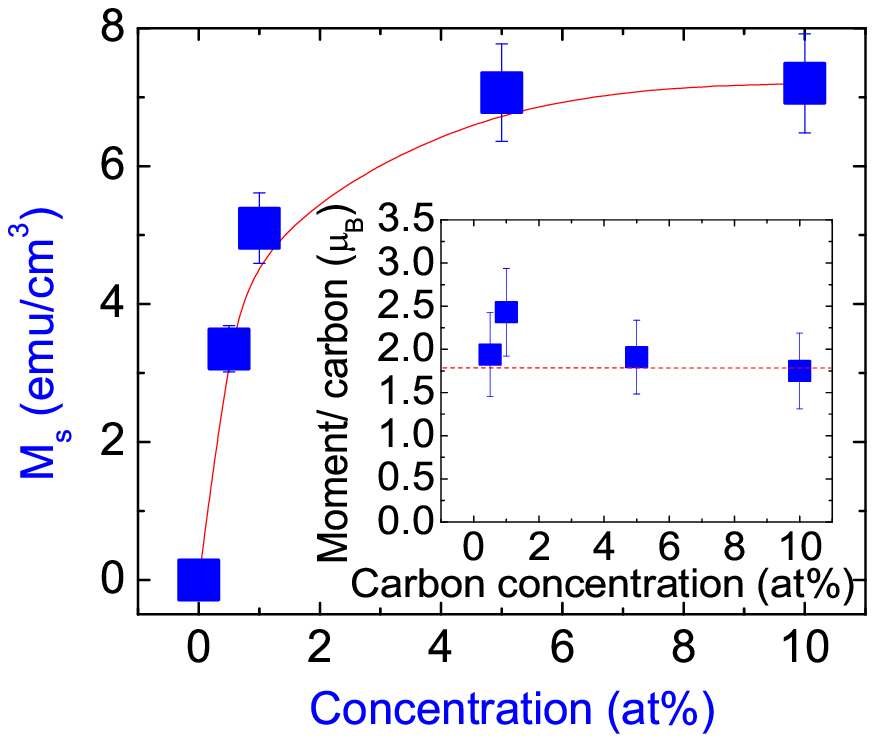} \\
Fig. 3 Pan et al.
\end{center}

\clearpage
\begin{center}
\includegraphics{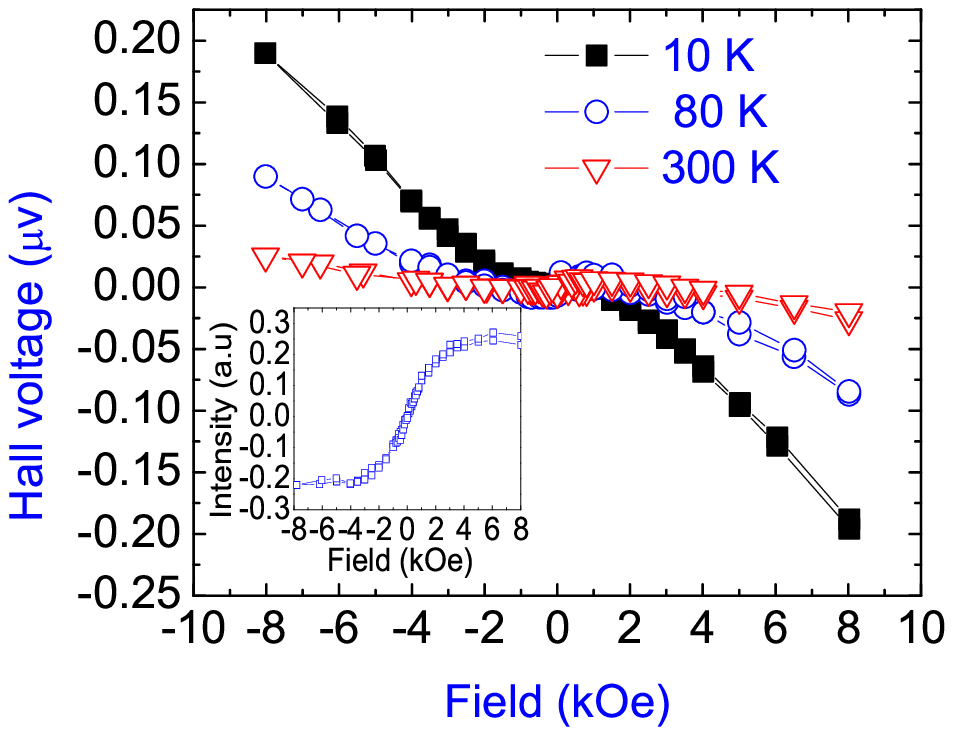} \\
Fig. 4 Pan et al.
\end{center}


\begin{thebibliography}{99}
\bibitem{Dietl} T. Dietl {\em et al.}, Science {\bf 287}, 1019 (2000).
\bibitem{Wu1}R. Q. Wu, {\em et al.}, Appl. Phys. Lett. {\bf 89}, 062505 (2006).
\bibitem{Park}M. S. Park and B. I. Min, Phys. Rev. B {\bf 68}, 224436 (2003).
\bibitem{Chien}C. H. Chien, {\em et al.}, J. Magn. Magn. Mater. {\bf 282}, 275 (2004).
\bibitem{Feng}X. Feng, J. Phys.: Condens. Matter {\bf 16}, 4251 (2004).
\bibitem{Ye}L. H. Ye, {\em et al.}, Phys. Rev. B {\bf 73}, 033203 (2006).
\bibitem{Buchholz}D. B. Buchholz, {\em et al.}, Appl. Phys. Lett. {\bf 87}, 082504 (2005).
\bibitem{Lee}J.-H. Lee, et al., MRS Spring Meeting, 2006.
\bibitem{Wu2} R. Q. Wu, {\em et al.}, Appl. Phys. Lett. {\bf 89}, 142501 (2006).
\bibitem{Coey} J. M. D. Coey, Solid State Sci. {\bf 7}, 660 (2005).
\bibitem{Venkatesan_PRL} M. Venkatesan {\em et al.}, Phys. Rev. Lett. {\bf 93}, 1772061 (2004).
\bibitem{Venkatesan_Nature} M. Venkatesan {\em et al.}, Nature (London) {\bf 430}, 630 (2004).
\bibitem{Makarova} T. L. Makarova {\em et al.}, Nature {\bf 413}, 716 (2001).
\bibitem{Kopelevich} Y. Kopelevich, {\em et al.}, J. Low Temp. Phys. {\bf 119}, 691 (2000).
\bibitem{Esquinazi} P. Esquinazi, {\em et al.}, Phys. Rev. B {\bf 66}, 024429 (2002).
\bibitem{Wang} X. Wang,  {\em et al.},  J. Phys.: Condens. Matter {\bf 14}, 10265 (2002).
\bibitem{Kopelevich03} Y. Kopelevich, {\em et al.}, Phys. Rev. B {\bf 68}, 092408 (2003).
\bibitem{Likodimos} V. Likodimos, {\em et al.}, Phys. Rev. B {\bf 72}, 045436 (2005).
\bibitem{Rode} A. V. Rode, {\em et al.},  Phys. Rev. B {\bf 70}, 054407 (2004).
\bibitem{Han} K. H. Han, {\em et al.}, Adv. Mater. {\bf 14}, 753 (2002).
\bibitem{Esquinazi03} P. Esquinazi ,  {\em et al.},  Phys. Rev. Lett. {\bf 91},  227201 (2003).
\bibitem{Esquinazi04} P. Esquinazi ,  {\em et al.},  Carbon {\bf 42},  1213 (2004).
\bibitem{Talapatra} S. Talapatra, {\em et al.}, Phys. Rev. Lett. {\bf 95},  097201 (2005).
\bibitem{Lehtinen} P. O. Lehtinen, {\em et al.}, Phys. Rev. B {\bf 69}, 155422 (2004).
\bibitem{Wu3} R. Q. Wu, {\em et al.}, Appl. Phys. Lett. {\bf 86}, 122510 (2005).
\bibitem{Payne} M. C. Payne {\em et al.}, Rev. Modern. Phys. {\bf 64}, 1045 (1992).
\bibitem{Milman} V. Milman {\em et al.}, J. Quant. Chem. {\bf 77}, 895 (2000).
\bibitem{Vanderbilt} D. Vanderblit, Phys. Rev. B  {\bf 41}, 7892 (1990).
\bibitem{Perdew} J. P. Perdew and A. Zunger, Phys. Rev. B  {\bf 23}, 5048 (1981).
\bibitem{Monkhorst} J. Monkhorst and J. Pack, Phys. Rev. B  {\bf 23}, 5188 (1976).
\bibitem{Madelung} O. Madelung {\em et al.}, Numerical Data and Functional Relationships in Science and Technology, (Springer-Verlag, Berlin, 1982), Vol. 17.
\bibitem{Kohan} A. F. Kohan {\em et al.}, Phys. Rev. B {\bf 62}, 15019 (2000).
\bibitem{Xu} Y. N. Xu and W. Y. Ching, Phys. Rev. B {\bf 48}, 4335 (1993).
\bibitem{Van} L. H. Van {\em et al.}, J. Alloy \& Comp. (in press). 
\bibitem{Ramqvist} L. Ramqvist {\em et al.}, J. Phys. Chem. Solids {\bf 30}, 1835 (1969).
\bibitem{Theodoropoulou} N. Theodoropoulou {\em et al.}, Phys. Rev. Lett. {\bf 89}, 107203 (2002).
\bibitem{akai} H. Akai, Phys. Rev. Lett. {\bf 81}, 3002 (1998).
\bibitem{sato} K. Sato, {\em et al.}, J. Phys.: Condens. Matter {\bf16}, S5491 (2004).

\end{thebibliography}
\end{document}